\documentclass[12pt]{iopart}
\usepackage{epsf}
\eqnobysec
\begin{document}
\title[Stratified disordered media]{Stratified disordered media: Exact solutions for transport parameters and their self-averaging properties}

\author{M. Clincy\dag\ and H. Kinzelbach\ddag}
\address{\dag\ Department of Physics and Astronomy, University of Edinburgh, Edinburgh EH9 3JZ, U.K.}
\address{\ddag\ Institut f\"ur Theoretische Physik, Philosophenweg 19, 69120 Heidelberg, Germany}

\ead{m.clincy@ed.ac.uk  kibach@tphys.uni-heidelberg.de}

\begin{abstract}
We investigate the transport of a passive tracer in a
two-dimensional stratified random medium with flow parallel and perpendicular 
to the strata. Assuming a Gaussian random flow with a Gaussian correlation
function, it is not only
possible to derive exact expressions for the temporal behaviour of the
dispersion
coefficients characterising the tracer transport, but also to
investigate the self-averaging
properties of these quantities.
We give explicit results for the dispersion coefficient and its mean
square fluctuations as a function of time. As it turns out, the sample
to sample
fluctuations which are encoded in the latter quantity remain finite
even in the asymptotic
limit of infinite times, which implies that the given two-dimensional
model is not
self-averaging.
\end{abstract}

\submitto{\JPA}

\maketitle

\section{Introduction}\label{intro}

20 years after the investigation of transport in stratified random
media by Matheron
and de Marsily \cite{Ma80}, their model is still intriguing. Not only
because it is exactly
solvable, but also as it serves as a prototype of models exhibiting
super-diffusive
behaviour \cite{Bo90b}.
As such it found applications in various fields like
hydromechanics \cite{Da87}, \cite{Sa91}
and polymer physics \cite{Do89}, \cite{Je00}.\\
The original model can be described as follows: Consider a
two-dimensional random medium consisting
of distinct parallel layers (or strata) which extend in
$x_1$-direction and have varying permeability.
The resulting velocity field $\mathbf{u}$ for a fluid flow in this
medium depends only on the
component $x_2$ transverse to the layers. In a random ensemble of such
stratified media, it is a
random field with given statistical properties. \\
The properties of such two-dimensional stratified models are well
known in the asymptotic limit
of large times, $t\to\infty$, and for the case of vanishing local
diffusion,
see e.g. \cite{Bo90a}, \cite{Re90}, \cite{Ra93a}.
The large scale transport behaviour is roughly characterised by the
mean square displacement
of the tracer particles, appropriately averaged over the ensemble of
the random flow fields.
The time derivative of this quantity encodes the large scale diffusion
or, as it is called,
`dispersion' coefficient $D(t)$ of the tracer cloud as a function of
time.
The transport is called diffusive if this coefficient becomes a finite
constant in the asymptotic
long-time limit $t \to \infty$, it is called super-diffusive if it
asymptotically grows with time
as $D(t)\sim t^\alpha$ with $\alpha >0$.
For a flow field parallel to the strata the model exhibits (for $t\to
\infty$)
such a super-diffusive behaviour with $\alpha =1/2$.
In the more general case, where one also has a transversal flow
component, the
dispersion coefficient eventually tends to a constant value, i.e. the
transport shows normal diffusive behaviour.\\
However, what is still lacking, to the best of our knowledge, is a
detailed analysis of the
pre-asymptotic behaviour of the model and, even more interestingly, of
the corresponding sample
to sample fluctuations of the dispersion coefficient. The latter is
crucial for the reliability
of the model.
Small fluctuations indicate that the transport parameters which are
constructed as averages over
the random ensemble are indeed also representative for the behaviour
of individual realizations
of the ensemble. \\
In the present article, we try to close this gap. Assuming a
stratified Gaussian random flow with a
Gaussian correlation function, we derive exact expressions for the
dispersion coefficient and
its mean square fluctuations as a function of time. The latter
quantity encodes the sample to
sample fluctuations of the dispersion coefficient. We show that it
remains finite even in the
asymptotic limit of infinite times $t\to\infty$ as long as the initial
tracer distribution is
of finite extent which implies that in the given two-dimensional model
the dispersion coefficient,
surprisingly, is not self-averaging. \\
The paper is organised as follows: In Section \ref{model} we give a brief
description of the two-dimensional
Matheron-de Marsily model and define the corresponding transport
parameters.
The next two sections deal with the exactly stratified model (i.e. a
model where the flow field is
aligned parallel to the strata), and with the more general case of a
non-vanishing transverse flow
component. In Section \ref{ave} we analyse the self-averaging properties of
the given transport parameters.

\section{The model}\label{model}

The transport of a passive tracer particle by a two-dimensional flow
field $\mathbf{u}(\mathbf{x})$
is modelled by a Langevin equation
${\rm d} \mathbf{x}(t)/{\rm d}t= \mathbf{u}(\mathbf{x}) + \mbox{\boldmath$\xi$\unboldmath}(t)$,
where $\mathbf{x}(t)$ denotes the position-vector of the particle, and
\boldmath$\xi$\unboldmath$(t)$
is a Gaussian white noise which generates the local
diffusion process. It has zero mean $\left<\xi_i(t)\right> = 0$ (with
$i=1,2$), and a
correlation function $\left<\xi_i(t)\xi_j(t')\right>= 2D_{ij}\,
\delta(t-t')$,
where $\left<...\right>$ stands for the average over the white noise
ensemble.
We assume a diagonal local diffusion tensor, 
$D_{ij}=\delta_{ij} \,D_i$ which,
however, may have different entries for the longitudinal and the
transversal diffusion constants, $D_1 \neq D_2$. \\
In a two-dimensional stratified medium the flow field $\mathbf{u}$
depends on $x_2$ only.
In the model investigated here, one has 
$\mathbf{u}(\mathbf{r})= (u_0+u'(x_2), v_0)$,
where $u_0$ and $v_0$ denote the components of a constant drift
velocity in $1$-and $2$-direction, whereas $u'(x_2)$ is the random flow 
contribution generated by the
layer to layer permeability variations (see e.g. \cite{Sa91}).\\
The Langevin equation therefore reduces to
\begin{eqnarray}
  \frac{\rm d}{{\rm d}t}\, x_1(t)  &= & u_0 + u'(x_2(t)) + \xi_1(t)\label{lang1}\\
  \frac{\rm d}{{\rm d}t}\, x_2(t)& = & v_0 + \xi_2(t)\label{lang2} 
\quad .
\end{eqnarray} 
The flow contribution $u'(x_2(t))$ is a random function chosen from an
appropriate
stationary ensemble which defines its stochastic properties. In the
model investigated here,
we assume that it has zero mean and a Gaussian correlation
function,
\begin{eqnarray}
  \overline{u'(x_2)} &=& 0 \label{u-average}\\
  \overline{u'(x_2)u'(x_2')} &=&
  \sigma_0^2 \, \exp{\left(-\frac{(x_2-x_2')^2}{2l^2}\right)} \label{u-corr}
\end{eqnarray}
The over-bar denotes the average with respect to disorder ensemble,
the parameter
$\sigma_0$ quantifies the disorder strength, and $l$ denotes the
correlation length
of the field.
As is shown below, these two relations uniquely determine the temporal
behaviour of the
dispersion coefficient, higher correlation functions do not enter.
This, however, is no longer true if one deals with the sample to
sample fluctuations of
this dispersion coefficient. In this case one needs also an explicit
expression for
the four-point function of the random field.
For simplicity, and without severe restriction of generality, let us
assume a Gaussian ensemble, so all higher correlation functions can be
decomposed
into products of the given two-point function.\\
Equation (\ref{lang1}) and (\ref{lang2}) define a system of
differential equations which is readily solved by
\begin{eqnarray}
  x_1(t)|_{\mathbf{x_0}} &=&x_{01} + u_0t + \int_0^t {\rm d}t' u'(x_2(t')) + \int_0^t
  {\rm d}t'\;\xi_1(t')\label{x1}\\
  x_2(t)|_{\mathbf{x_0}} & = & x_{02} + v_0t +\int_0^t {\rm d}t'\;\xi_2(t')\label{x2} \quad . 
\label{solution}
\end{eqnarray}
The single tracer particle subject to a given noise history is characterized by 
its trajectory $\mathbf{x}(t)|_{\mathbf{x_0}}$, where the subscript $\mathbf{x_0}$
indicates the fact that the tracer coordinates at a given time $t$ still depend on 
the initial injection position 
$\mathbf{x}(t=0)|_{\mathbf{x_0}} \equiv {\mathbf{x_0}}=(x_{01},x_{02})$.\\
In the general case where one has a spatially extended injection of material at 
time $t=0$ the injection points $\mathbf{x_0}$ are distributed according to a 
distribution $\rho({\mathbf{x_0}})$ which without restriction is chosen to be normalised,  
\begin{equation}
\int d^2 x_0 \;\, \rho(\mathbf{x_0})=1\label{init_dist}.
\end{equation}
The shape of the resulting concentration distribution is characterised by its moments 
\begin{equation}
\left< {x_i(t)^n} \right> \equiv \int d^2 x_0  \;\, \rho(\mathbf{x_0}) \; 
\left< {x_i(t)|_{\mathbf{x_0}}^n}\right> 
\label{moments}
\end{equation} 
(with $n=1,2,\dots$) which in particular are used to define the centre-of-mass velocity and 
the dispersion coefficients of the resulting concentration plume. 
The centre-of-mass velocity (for a given realization of the flow field) reads 
\begin{equation}
  u_i(t)=\frac{{\rm d}}{{\rm d}t}\; \left<x_i(t) \right> \quad .   \label{cmveloc}
\end{equation}
The dispersion coefficients which give the spreading rate of the plume are defined by 
\begin{equation}
  D_{ij}(t) = \frac{1}{2} \, \frac{{\rm d}}{{\rm d}t} \;
  \Big\{\,{ \left<{x_i(t) \, x_j(t)}\right> - 
          \left<{x_i(t)}\right> \, \left<{x_j(t)}\right> \; }\Big\} \label{dispcoeff}
\end{equation}
In a stochastic approach, the given single medium is viewed as 
one particular realization of a spatial stochastic process. Large scale transport
coefficients are derived from appropriately constructed averages over the ensemble
of all possible medium realizations. These averages represent statistical properties
of the (artificial) ensemble as a whole. At first glance, they therefore seem to 
be of limited predictive value with respect to the properties of a single given
aquifer. For appropriately chosen quantities, however, in many cases the fluctuations 
from realization to realization become small as soon as the cloud has sampled a
sufficiently large representative part of the given medium. 
As soon as the transport properties found in different realizations of the
medium then fluctuate only weakly around the corresponding ensemble averages, these 
averages again also predict properties characteristic of a single typical realization 
of the medium. \\
In what follows, we investigate such ensemble averages of the centre-of-mass velocity and 
the dispersion coefficients. We call the averaged quantities corresponding 
to (\ref{cmveloc}) and (\ref{dispcoeff}) ``effective'' quantities. 
The resulting effective centre-of-mass velocity and  effective dispersion coefficient 
are defined as 
\begin{equation}
  u_i^{\rm eff}(t) \;\equiv\; \overline{u_i(t)} \;=\; \frac{{\rm d}}{{\rm d}t}\; 
  \overline{\left<x_i(t) \right>}
\end{equation}
and 
\begin{equation}
  D_{ii}^{\rm eff}(t) \;\equiv\; \overline{ D_{ii}(t)} \;=\;\; 
 \frac{1}{2} \, \frac{{\rm d}}{{\rm d}t} 
  \left\{\, \overline{\left<x_i^2(t)\right>} -\overline{\,
      \left<x_i(t)\right>^2\,} \; \right\} \quad . 
\label{deff_pt} 
\end{equation}
The non-diagonal dispersion coefficients vanish due to symmetry reasons. 
Note that two types of averages are involved, the average over the white noise 
which generates the local diffusion process, indicated by the brackets, and 
the average over the disorder ensemble, indicated by the over-bar. The order 
in which these averages are performed is crucial, as the following 
slightly different definition of an averaged dispersion coefficient demonstrates. 
We define what we call ``ensemble dispersion 
coefficients'' as 
\begin{eqnarray}
  D_{ii}^{\rm ens}(t) &=& \frac{1}{2}\, \frac{{\rm d}}{{\rm d}t} \;
  \left\{\, \overline{\left<x_i^2(t)\right>} -\overline{\,
      \left<x_i(t)\right> \,}^{\; 2} \, \right\} \quad . 
  \label{dens_pt}
\end{eqnarray}
This quantity, which also is well known and frequently used in the literature, 
represents the dispersion characteristics of the whole ensemble of realizations of the flow field. 
It takes into account an artificial dispersion effect caused by fluctuations of the centre-of-mass 
positions of the tracer distributions in different realizations of the inhomogeneous medium. 
This effect is suppressed in the effective dispersion coefficient as defined in (\ref{deff_pt}) 
because there the centre-of-mass positions are superimposed before the ensemble average is 
performed. 
For an extensive discussion of these problems and the related literature, see \cite{De01a} 
and \cite{De01b}. 
In general, the experimentally observable dispersion which is a property related to a single given 
realization is represented by the effective quantity $D^{\rm eff}(t)$. 
Usually, at finite times the ensemble quantity overestimates the experimentally 
observable dispersion considerably, see again the detailed discussion 
given in \cite{De01a}, \cite{De01b}.
In the asymptotic long-time limit $t\to\infty$, however, one usually expects 
that the two types of dispersion coefficients become equal.\\ 
For the model investigated here, the full temporal behaviour of the given transport coefficients 
can be evaluated explicitly. The corresponding calculations are elementary but somewhat 
tedious. They are summarized in \ref{app1}. Using the explicit solution (\ref{solution}) and 
performing the appropriate averages, one finds the more or less trivial results 
\begin{equation}
u_{1}^{\rm eff}(t)= u_0 \;\; , \;\; u_{2}^{\rm eff}(t)=  v_0 \qquad \mbox{and} \qquad
D_{22}^{\rm ens}(t) = D_{22}^{\rm eff}(t) =  D_2  
\label{triv}
\end{equation} 
valid for all times. 
A more interesting behaviour follows for the longitudinal dispersion coefficients. Here 
one derives the relations  
\begin{eqnarray}
D_{11}^{\rm ens}(t)&=& D_1 + \frac{1}{2}\partial_t \; 
\int\limits_{k}\int\limits_{k'}\overline{\tilde{u}(k)\tilde{u}(k')}\times\nonumber\\
& & \quad\times\int_0^t {\rm d}t'\int_0^t {\rm d}t'' {\rm e}^{-(k^2 D_2 t' +k'^2 D_2 t''+2 k k' D_2 \min(t',t''))}
\quad\qquad \label{D11ens}
\\
D_{11}^{\rm eff}(t)&=& D_{11}^{\rm ens}(t) - 
\frac{1}{2} \partial_t \;  \overline{\left(\int_k\tilde{u}(k)
\int \! d^2x_0 \, \rho(\mathbf{x}_0){\rm e}^{-{\rm i}k\,y_0}\int_0^t {\rm d}t' {\rm e}^{-k^2 D_2 t'}\right)^2\;} \qquad 
\label{D11eff}
\end{eqnarray}
where we use  $\int_{k}\dots \equiv (2\pi)^{-2}\int {\rm d}^2 k \dots$ as a convenient short-hand 
notation.\\
One can already observe three simple features: First of all, the effective and the ensemble 
dispersion coefficient in the direction transversal to the flow field are both equal to the local 
dispersion coefficient $D_2$. The heterogenities of the model do not change this property 
because they do not depend on the fluctuating part of the flow field. 
The particles perform a simple random walk in $x_2$-direction. 
Secondly, the effective dispersion coefficient in $x_1$-direction does not depend on the 
longitudinal extension of the initial distribution. This stems from the infinite correlation 
length of the flow field in that direction. 
The third observation is that the ensemble dispersion coefficient does not show any dependence on 
the size of the initial distribution.\\
What remains to be evaluated is the explicit behaviour of the longitudinal dispersion 
coefficient. 
In the following we will discuss the cases $v_0=0$, where the fluid
flow is aligned with the strata of the medium (the so called `exactly stratified model'), 
and $v_0\neq 0$ separately as they give rise to a different asymptotic behaviour. 
We will further distinguish between a point-like initial distribution at $x_{10}=x_{20}=0$ 
(which implies $\rho({\mathbf x_0}) = \delta \left({\mathbf x_0}\right)$ in (\ref{moments})) and 
an spatially extended distribution
\begin{eqnarray}
  \rho({\mathbf x_0}) =
  \frac{1}{\sqrt{2\pi}L}\exp{\left(-\frac{x_{02}^2}{2L^2}\right)}\delta\left(x_{01}\right) 
\label{InitialDistrib}
\end{eqnarray} 
where the injection points are aligned along a line in $x_2$ direction. As it turns out, the line shape is no restriction of generality since a finite extent in $x_1$-direction does not
alter the given results.

\section{The exactly stratified medium}\label{strat}

\begin{figure}
\begin{center}
\epsfbox{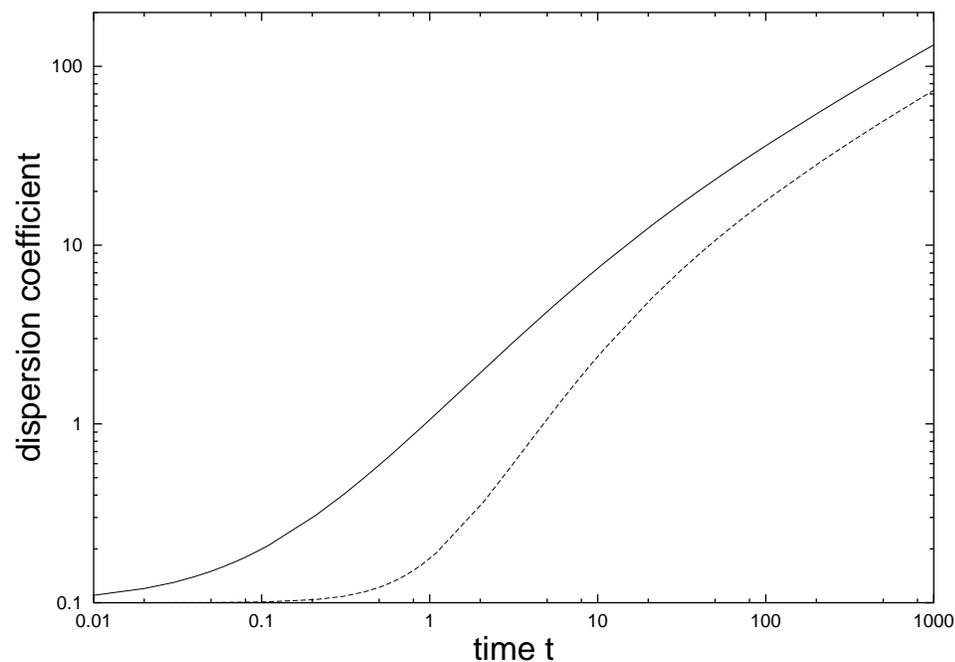}
\end{center}
\caption{\label{dkpkt}Ensemble (\full) and effective (\broken) dispersion coefficients of an exactly stratified medium for a point source; $\tau_D = 10$, $D_1 = 0.1$, $\sigma_0^2 = 1$}
\end{figure}

\begin{figure}
\begin{center}
\epsfbox{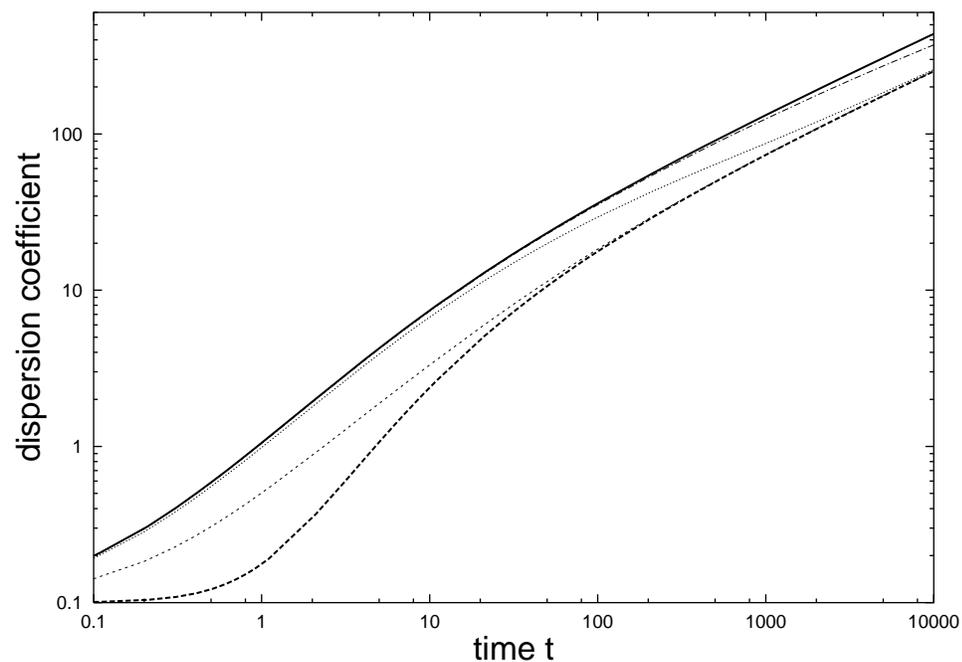}
\end{center}
\caption{\label{dklin}Ensemble (\full) and effective dispersion coefficients of an exactly stratified medium for a point source (\broken) and line sources of various length ($L=1$ (\dashed), $L=10$(\dotted), $L=100$(\chain)) ; $\tau_D = 10$, $D_1 = D_2 = 0.1$, $\sigma_0^2 = 1$ }
\end{figure}

For $v_0=0$ and a point-like initial distribution the expressions for the effective
dispersion coefficients given by equations (\ref{D11ens}) and (\ref{D11eff}) 
can be evaluated easily. One derives   
\begin{eqnarray} 
  D_{11}^{{\rm ens}}(t) &=& D_1 + \sigma_0^2 \; \tau_D
  \left(\sqrt{1+\frac{2t}{\tau_D}}-1\right)
\label{dens_pt_v0}\\
D_{11}^{{\rm eff}}(t) &=& D^{{\rm ens}}(t) - \sigma_0^2 \; \tau_D
\left(\sqrt{1+\frac{4t}{\tau_D}} \,-\, \sqrt{1+\frac{2t}{\tau_D}}\;
\right) \label{deff_pt_v0}
\end{eqnarray}
where the time-scale $\tau_D=l^2/D_2$ is a measure for the diffusive transport over one 
transversal correlation length. 
The result is plotted in Figure \ref{dkpkt}. 
There are two characteristic regimes. 
For $t\ll\tau_D$ the dispersion coefficients asymptotically are given as
\begin{eqnarray}
  D_{11}^{{\rm ens}}(t) & = & D_1 + \sigma_0^2 t + O\left(\left(t/\tau_D\right)^2\right) 
\label{3.28}\\
  D_{11}^{{\rm eff}}(t)& = & D_1 +
  O\left(\left(t/\tau_D\right)^2\right)\label{3.29}
\end{eqnarray}
In this regime there exist great fluctuations between the centre of mass
coordinates of the source in different realizations of the flow field,
the transport is dominated by advection which explains the linear time
dependence of the ensemble
dispersion coefficient. In each realization the particles only
experience local diffusion which can be
seen in the effective value which is approximately constant. This
behaviour characterises the Taylor regime
as already described by \cite{Da87}.\\
In the opposite long time limit, $t \gg \tau_D$, the dispersion coefficients take
the following asymptotic form
\begin{eqnarray}
  D_{11}^{{\rm ens}}(t) & = & D_1 + \sigma_0^2 \, \sqrt{2\tau_D} \; \;
  t^{1/2} \;+ O\left(\left(t/\tau_D\right)^{-1/2}\right)
\label{3.30}\\
D_{11}^{{\rm eff}}(t)& = & D_1 + \sigma_0^2 \, [2-\sqrt{2}] \, \sqrt{2\tau_D}  \;\; t^{1/2}\;
 \;+ O\left(\left(t/\tau_D\right)^{-1/2}\right)
\label{3.31}
\end{eqnarray}
They show the well known anomalous diffusion $t^{1/2}$-behaviour with 
different pre-factors for the two types of dispersion coefficients, a 
result which is already described in \cite{Bo90b}, \cite{Do89} and \cite{Re90}. 
These authors derive this result by considering the asymptotic behaviour of a 
particle exhibiting a random walk and experiencing a different velocity after each time 
step, i.e. in each layer. 
The factor of $\left(2-\sqrt{2}\right)$ between $D_{11}^{{\rm ens}}(t)$ and 
$D_{11}^{{\rm eff}}(t)$ is due to large fluctuations in the centre of mass coordinates of the source distribution in different realizations.\\
For an extended initial distribution (in particular for the line source given by 
equation (\ref{InitialDistrib}) the expression for the ensemble coefficient
$D_{11}^{{\rm ens}}(t)$ as given in (\ref{dens_pt_v0}) remains
unchanged. 
The effective dispersion coefficient, however, now reads
\begin{eqnarray}
&&  D_{11}^{{\rm eff}}(t)|_{\rm line} =  D^{{\rm ens}}(t) -\nonumber \\
 & & \quad - \sigma_0^2\sqrt{\tau_D}\sqrt{\tau_D+\tau_L} \left\{
    \sqrt{1+\frac{4t}{\tau_D+\tau_L}}-\sqrt{1+\frac{2t}{\tau_D+\tau_L}}
  \right\}
\label{deff_lin_v0}
\end{eqnarray}
Another time-scale that becomes relevant, $\tau_L=L^2/D_2$, is the
typical time for diffusive 
transport over the width $L$ of the initial concentration distribution. 
For times much smaller than $\tau_L$ the effect of a finite initial 
extent of the injection region is equivalent to averaging over various 
realizations. The dispersion coefficient tends
to the ensemble value given by given by equation (\ref{dens_pt_v0}).
For times much longer than $\tau_L$ the system behaviour no longer 
depends on the initial width of the injection region. The effective 
dispersion coefficient, therefore, tends to 
the value derived for a point source, equation (\ref{deff_pt_v0}). 
The resulting temporal behaviour is plotted in Figure \ref{dklin}. For an infinite $L$ the value of $D_{11}^{{\rm eff}}(t)|_{\rm line}$ tends
to $D_{11}^{{\rm ens}}(t)$ for all
times.

\section{The general case}\label{gen}

\begin{figure}
\begin{center}
\epsfbox{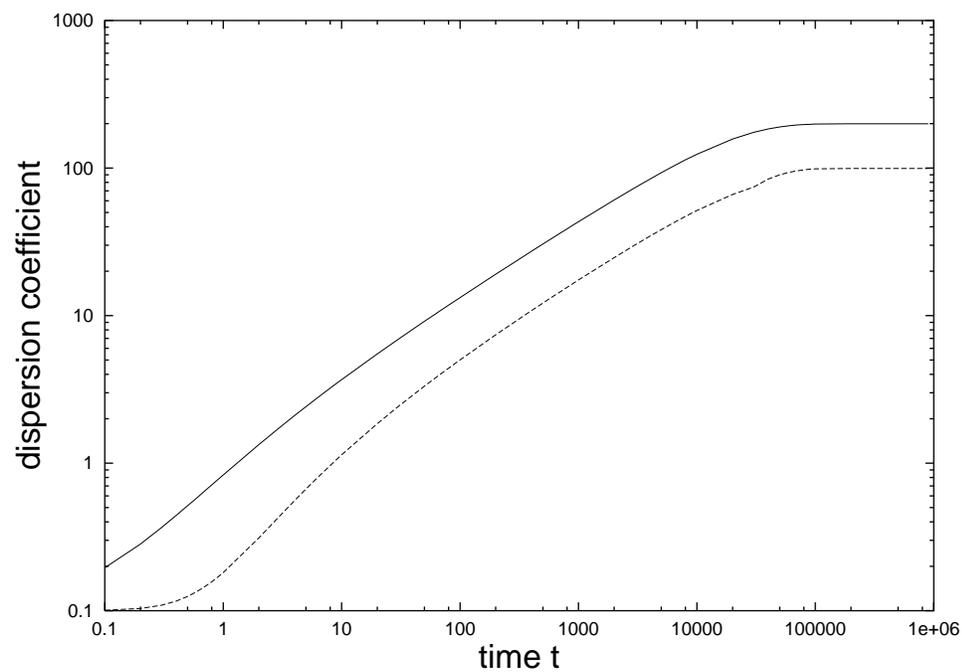}
\end{center}
\caption{\label{dku2}Ensemble (\full) and effective (\broken) dispersion coefficients for a stratified medium with a flow component transverse to the strata; $\tau_D = 1$, $\tau_v = 80$, $D_1 = 0.1$, $\sigma_0^2 = 1$}
\end{figure}

Unlike the exactly stratified model, the long-time behaviour for the
case $v_0\neq 0$ shows normal diffusion for asymptotically large times, 
$t\to \infty$. Evaluating equations (\ref{D11ens}) and (\ref{D11eff}) 
now lead to the more complicated expressions 
\begin{eqnarray}
&& D_{11}^{{\rm ens}}(t) = D_1 + \sqrt{\frac{\pi}{2}}\, \sigma_0^2 \, \tau_v 
\left[
   {\rm erf}\left( \sqrt{\frac{t^2/2\tau_v^2}{1+2t/\tau_D} } \; \right) \; + 
\right. 
\nonumber \\
&& \quad + \; \exp\left( \frac{\tau_D^2}{2\tau_v^2}\right) 
\left\{ {\rm erf}\left( \sqrt{\frac{t^2/2\tau_v^2}{1+2t/\tau_D} } \;(\tau_D/t+1)\right) 
- \left. {\rm erf}\left(\sqrt{\frac{\tau_D^2}{2\tau_v^2}}\right)\right\}\right] 
\qquad 
\label{dens_pt_v}
\\
&& D_{11}^{{\rm eff}}(t)  = D^{{\rm ens}}(t) - \sqrt{\frac{\pi}{2}}\, \sigma_0^2 \, \tau_v
\left[ \;
    {\rm erf}\left( \sqrt{\frac{t^2/2\tau_v^2}{1+2t/\tau_D} } \; \right)   \; + 
\right.
\nonumber\\
&& \qquad + \; \left. \exp\left(  \frac{\tau_D^2}{2\tau_v^2} (1+4t/\tau_D) \right)
\left\{ 
   {\rm erf}\left(\sqrt{\frac{t^2/2\tau_v^2}{1+2t/\tau_D}} \; (\tau_D/t +3) \right) \; - 
\right.\right.
\nonumber\\
&& \qquad \qquad 
-  \left.\left.  {\rm erf}\left(\sqrt{ \frac{\tau_D^2}{2\tau_v^2} (1+4t/\tau_D)}\;
\right)\; \right\} \; \right] 
\qquad 
\label{deff_pt_v}
\end{eqnarray} 
where ${\rm erf}(x)$ is the Gaussian error function as defined in \cite{Ab84}. The time-scale $\tau_v=l/v_0$ is set by the time necessary to 
transport a tracer particle advectively over one correlation length of the flow field.\\
For $\tau_D \ll \tau_v$ three regimes can be distinguished which is shown in Figure \ref{dku2}. For small times the advection has only a negligible influence on the dispersion coefficient. We can therefore identify the two superdiffusive regimes from the exactly stratified case, i.e. the Taylor-regime given by eqns. (\ref{3.28}) and (\ref{3.29}) for $t\ll\tau_D\ll\tau_v$ (neglecting terms of the order $O\left(\tau_D/\tau_v\right)$ and $O\left(\left(t/\tau_v\right)^2\right)$, resp.) and the superdiffusive $t^{1/2}$-behaviour given by eqns. (\ref{3.30}) and (\ref{3.31}) for $\tau_D\ll t\ll\tau_v$ (neglecting terms of the order $O\left(\tau_D/t\right)$ and $O\left(\left(t\tau_D/\tau_v^2\right)^2\right)$ , respectively). In the asymptotic limit $t\to\infty$ the system exhibits normal diffusion. The values of the dispersion coefficients are 
\begin{eqnarray}
  D_{11}^{{\rm ens}}(t)&=& D_1 + \sqrt{2\pi}\; \sigma_0^2 \, \tau_v +
  O\left(\tau_D/\tau_v\right) + O\left(\tau_v/\sqrt{t\tau_D}\right)
\label{4.14}\\
D_{11}^{{\rm eff}}(t)&=& D_1 + \frac{1}{2} \, \sqrt{2\pi} \;
\sigma_0^2 \, \tau_v + O\left(\tau_D/\tau_v\right) +
O\left(\tau_v/\sqrt{t\tau_D}\right)
\label{4.15}
\end{eqnarray}
which shows that, somewhat unexpectedly, the two quantities do not become equal even in the normal-diffusive case. This model seems to be a counter-example to Metzger's conjecture \cite{Me99} that for normally diffusive models the effective and the ensemble dispersion coefficients should always become equal in the asymptotic limit.\\
One remarkable feature of the cross-over between the superdiffusive and the normal diffusive regime is that it does not occur at $\tau_v$, but at a time scale $\tau_v^2/\tau_D = D_2/v_0^2$ which is independent of the correlation length of the system.\\
For the case $\tau_v \ll\tau_D$ the behaviour is much simpler. Advection dominates over diffusion, i.e. the diffusive time scale $\tau_D$ cannot be resolved. Therefore there are only two regimes: for $t \ll\tau_v$ we again find the Taylor regime with a linear growth of the ensemble dispersion coefficient, at $\tau_v$ there is a cross-over to a normal diffusive regime given by
\begin{eqnarray}
D_{11}^{\rm ens}(t)&=& D_1 + \sqrt{\frac{\pi}{2}}\sigma_0^2\tau_v +
O\left(\frac{\tau_D}{t}\right) + O\left(\sqrt{\frac{\tau_v}{\tau_D}}\right) \label{4.14a}
\end{eqnarray} 
Because the diffusive effect on the mixing is suppressed by advection, the effective dispersion coefficient is equal to $D_1$ in leading order:
\begin{eqnarray}
D_{11}^{\rm ens}(t)&=& D_1 + O\left(\frac{\tau_D}{t}\right) + O\left(\sqrt{\frac{\tau_v}{\tau_D}}\right)
\end{eqnarray}

\section{Self-Averaging}\label{ave}

\begin{figure}
\begin{center}
\epsfbox{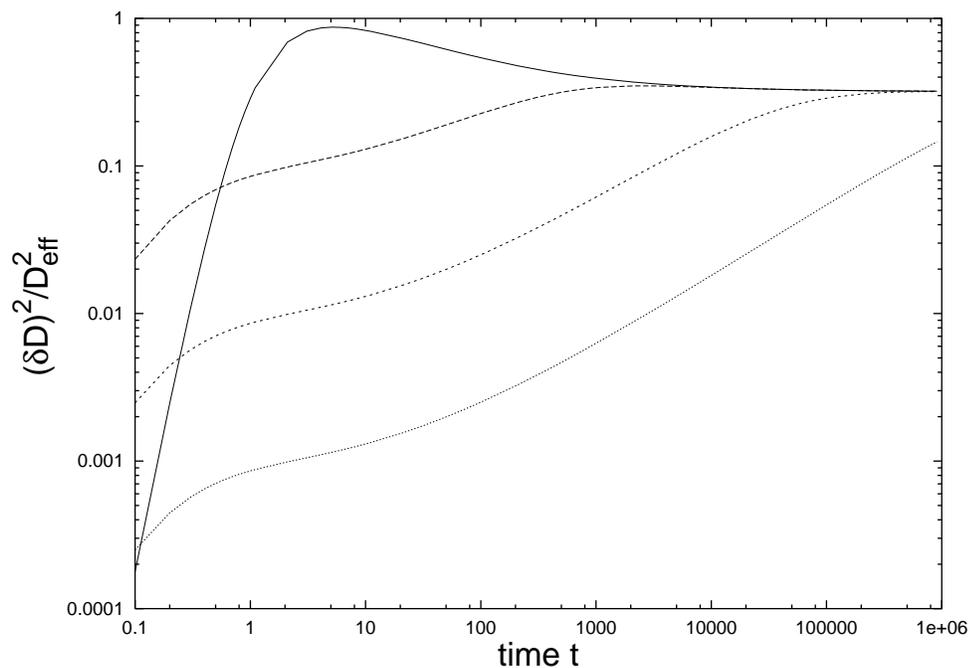}
\end{center}
\caption{\label{fluk}Normalised fluctuations of the dispersion coefficient in an exactly stratified medium for a point source (\full) and line sources with $L = 10$ (\broken), $L=100$ (\dashed), $L=1000$ (\dotted); $\tau_D = 10$, $D_1 = D_2 = 0.1$, $\sigma_0^2 = 1$ }
\end{figure}

The question remains as to whether the effective ensemble-averaged parameters represent the
typical behaviour in a single realization of the statistical ensemble. 
To investigate this problem further, we calculate the root-mean-square
sample-to-sample fluctuations of the centre-of-mass velocity and the dispersion
coefficient which are given by 
\begin{eqnarray}
  \left( \delta u(t)\right)^2 &\equiv& \overline{\, u(t)^2 \,} -\overline{\, u(t)\,}^2\\
  \left( \delta D(t)\right)^2 &\equiv& \overline{\, D(t)^2 \,}
  -\overline{\, D(t)\,}^2 
\label{deltaD}
\end{eqnarray}
where $u(t)$ and $D(t)$ stand for the centre-of-mass velocity and the
dispersion coefficient
of the single realizations as defined in (\ref{cmveloc}) and (\ref{dispcoeff}). \\
Again, these quantities can be evaluated explicitly in the given model. 
As expected, we find that the effective velocity becomes 
self-averaging in the asymptotic long-time limit, 
$\delta u(t)\to 0$ for $t\to\infty$. 
In the opposite short-time case, $t\to 0$, the fluctuations are determined by the
fluctuation strength of the disorder $\sigma_0^2$. 
As an example, consider the simple case of 
an exactly stratified medium, $v_0=0$, with point-like injection (i.e. $L=0$). 
The explicit temporal behaviour of the root-mean-square fluctuations is found  
to be 
\begin{equation}
  \frac{(\delta u(t))^2}{\overline{u(t)}^2} \;  = \; \frac{(\delta u(t))^2}{{u_0}^2}\; = \;
  \frac{\sigma_0^2}{u_0^2}\; \sqrt{\frac{1}{1+4t/\tau_D}} \quad .
\label{3.47}
\end{equation}
Surprisingly, the situation is different in the case of the dispersion coefficient. 
It does not become self-averaging even in the asymptotic long-time limit. 
Here, in the limit  $t\to\infty$ the normalized fluctuations 
$\delta D/\overline{D}^2$ tend to a constant finite value 
for the exactly stratified case $v_0=0$ as well as for the case $v_0\neq 0$ where 
on has normal diffusion in this limit. \\
The explicit results are somewhat cumbersome and not very illustrative. 
The derivation and the explicit formulae are, therefore, deferred to 
\ref{app2}. 
Figure \ref{fluk} illustrates the qualitative behaviour for the case of an 
exactly stratified medium (i.e. $v_0=0$) and line injections of 
different lengths $L$. 
As already described in the previous two sections, the short time regime is 
characterized by an effective dispersion coefficient which is mainly determined 
by local diffusion. This quantity does not vary between different realizations of the 
flow field, the fluctuations therefore obviously vanish for $t\to 0$. 
For $t\to\infty$ the sample-to-sample fluctuations tend to a constant value, irrespective 
of the initial length of the injection source. The time-scale on which this asymptotic 
value is reached is given by $\tau_D +\tau_L$, i.e. it is larger for larger $L$. 
Consequently, an infinitely long source will show no fluctuations at all as all of the 
fluctuations of the flow field are already experienced right from the start.\\
The non-vanishing sample-to-sample fluctuations demonstrate 
the fact that the effective dispersion coefficient as defined by (\ref{deff_pt}) 
is  not a self-averaging quantity, it therefore does not 
represent the typical behaviour expected in a single representation of the flow field 
in the two-dimensional stratified model.

\appendix

\section{}\label{app1}

In the following appendix, we sketch the basic steps used to derive the results 
for the temporal behaviour of the transport coefficients discussed in the text. 
Starting point is the solution of the Langevin equation given by (\ref{x1}) 
and (\ref{x2}) which is inserted into equation (\ref{moments}) with $n=1$ and $n=2$. 
To perform the necessary white-noise averages, it is useful to replace the 
flow field by its Fourier representation, 
$u'(x_2(t))=\int_k {\tilde u'}(k)\;{\rm e}^{-{\rm i}k\,x_2(t)}$. 
After inserting (\ref{x1}) and (\ref{x2}), the resulting white-noise averages can 
be calculated using the general formula 
$ \left<\exp\{-{\rm i}\sum_j \int \! {\rm d}t' L_j(t')\,\xi_j(t')\}\right> = 
\exp\{-\sum_{j,l} \int \! {\rm d}t' L_j(t')\,D_{jl}\,L_l(t')\}$ 
and an appropriate choice of the arbitrary auxiliary function $L_j(t)$.
One ends up with 
\begin{eqnarray}
<x_1(t)|_{\mathbf{x_0}}> &=& x_{01} + u_0t + \int_0^t {\rm d}t'\int_{k} 
\tilde{u}'(k){\rm e}^{{\rm i}k\left(x_{02} + v_0t'\right)}{\rm e}^{-k^2D_2t'}
\label{x_1_0a}\\
<x_1^2(t)|_{\mathbf{x_0}}> &=& 2D_1t+\left(x_{01} + u_0t\right)^2+
\nonumber\\
&+& \: 2\left(x_{01} + u_0t\right)\int_0^t {\rm d}t'\int_{k} \tilde{u}'(k) {\rm e}^{{\rm i}k\left(x_{02} + v_0t'\right)}
{\rm e}^{-k^2D_2t'}+
\nonumber\\
&+& 2\int_0^t {\rm d}t'\int_0^{t'} {\rm d}t''\int_{k} \int_{k'} \tilde{u}'(k)\tilde{u}'(k')
{\rm e}^{-{\rm i}k\left(x_{02} + v_0t'\right)-{\rm i}k'\left(x_{02} + v_0t''\right)}\times\nonumber\\
& & \quad\times \exp\left(-k^2D_2t'-k'^2D_2t''-2kk'D_2\min(t',t'')\right)\quad 
\label{x_2_0a}
\end{eqnarray}
and 
\begin{eqnarray}
<x_2(t)|_{\mathbf{x_0}}> &=& x_{02} + v_0t 
\label{y_1_0}\\
<x_2^2(t)|_{\mathbf{x_0}}> &=& 2D_2t+\left(x_{02} + v_0t\right)^2 \quad .
\label{y_2_0}
\end{eqnarray}
These expressions still depend on the initial positions ${\mathbf x_0}$ of the particles. 
According to (\ref{moments}) one finally performs the remaining average over the initial 
distribution $\rho({\mathbf x_0})$. Inserted into the definitions for the effective and ensemble 
dispersion coefficients (\ref{deff_pt}) and (\ref{dens_pt}), and performing the 
ensemble average over the disordered flow field, this eventually leads 
to (\ref{triv}), (\ref{D11ens}), and (\ref{D11eff}).  
Using the explicit form for the initial distribution of equation (\ref{InitialDistrib}) and 
the expression for the disorder averages (\ref{u-average}) and (\ref{u-corr}) 
then yields the final explicit results given in the text. 

\section{}\label{app2}
In  \ref{app2}, we sketch the basic steps used to derive the results 
for the sample-to-sample fluctuations of the transport coefficients given in the text. \\
For an exactly stratified medium (i.e. $v_0=0$) and a  point-like injection 
(i.e. $L=0$), using the steps described in \ref{app1}, 
one derives the following results for the square of the dispersion 
coefficient $D(t)$ in a given realization of the 
medium as defined by (\ref{dispcoeff}) 
\begin{eqnarray}
  D(t)^2 &= &D_1^2\; + \\
  &+ & 2 D_1\int_{k_1}\int_{k_2}\tilde{u}(k_1)\tilde{u}(k_2)\int_0^t
  {\rm d}t_1
  {\rm e}^{-k_1^2D_2t-k_2^2D_2t_1}\left({\rm e}^{-2k_1k_2D_2t_1}-1\right)+\nonumber\\
  &+ &
  \int_{k_1}\int_{k_2}\int_{k_3}\int_{k_4}\tilde{u}(k_1)\tilde{u}(k_2)\tilde{u}(k_3)\tilde{u}(k_4)
  \times \nonumber\\
  & & \times
  {\rm e}^{-k_1^2D_2t-k_2^2D_2t_1}\left({\rm e}^{-2k_1k_2D_2t_1}-1\right){\rm e}^{-k_3^2D_2t-k_4^2D_2
    t_2}\left({\rm e}^{-2k_3k_4D_2 t_2}-1\right) \nonumber
\end{eqnarray} 
Assuming a Gaussian ensemble for the flow field, the four point function
$\overline{\tilde{u}(k_1)\tilde{u}(k_2)\tilde{u}(k_3)\tilde{u}(k_4)}$ 
factorizes as usual into products of two-point correlation functions. 
Using this decomposition, and a rescaling of the time integrations of the form 
$\tau_1:=t_i/t$ we get for the  root-mean-square sample-to-sample fluctuations  
of the dispersion coefficient defined by (\ref{deltaD})
\begin{eqnarray}
(\delta D(t))^2 &=& \frac{1}{2}\sigma_0^4 \tau_D t 
\int\limits_0^1 d\tau_1 \int\limits_0^1 d\tau_2 
\left\{ A\left(\frac{\tau_D}{2t},\tau_1,\tau_2\right)^{-1/2}+ 
B\left(\frac{\tau_D}{2t},\tau_1,\tau_2\right)^{-1/2} \right. 
\nonumber\\
&& \hspace*{-2.5cm} 
- 2 \left[A\left(\frac{\tau_D}{2t},\tau_1,\tau_2\right)-\tau_2^2\right]^{-1/2} 
-2 \left[ B\left(\frac{\tau_D}{2t},\tau_1,\tau_2\right)-\tau_2^2\right]^{-1/2} 
\nonumber\\
&& \hspace*{-2.5cm} 
+ \left[A\left(\frac{\tau_D}{2t},\tau_1,\tau_2\right)-\left(\tau_1+\tau_2\right)^2\right]^{-1/2}
+ \left. \left[ B\left(\frac{\tau_D}{2t},\tau_1,\tau_2\right)-\left(\tau_1+\tau_2\right)^2\right]^{-1/2}
\right\} \qquad 
\label{3.48}
\end{eqnarray}
with
\begin{eqnarray}
A\left(\frac{\tau_D}{2t},\tau_1,\tau_2\right)&:= &\left(\frac{\tau_D}{2t}+2\right)\left(\frac{\tau_D}{2t}+\tau_1+\tau_2\right)\nonumber\\
B\left(\frac{\tau_D}{2t},\tau_1,\tau_2\right)&:= &\left(\frac{\tau_D}{2t}+1+\tau_1\right)\left(\frac{\tau_D}{2t}+1+\tau_2\right)\label{3.49}
\end{eqnarray} 
The calculation is easily generalized to the case of an 
extended injection (i.e. $L\neq 0$). There, the functions 
$A\left(\frac{\tau_D}{2t},\tau_1,\tau_2\right)$ and 
$B\left(\frac{\tau_D}{2t},\tau_1,\tau_2\right)$ are replaced by 
\begin{eqnarray}
A\left(\frac{\tau_D+\tau_L}{2t},\tau_1,\tau_2\right)&:=& (\frac{\tau_D + \tau_L}{2t}+2)(\frac{\tau_D + \tau_L}{2t}+\tau_1+\tau_2)\nonumber\\
B\left(\frac{\tau_D+\tau_L}{2t},\tau_1,\tau_2\right)&:=& (\frac{\tau_D+\tau_L}{2t}+1+\tau_1)(\frac{\tau_D+\tau_L}{2t}+1+\tau_2)\label{3.53}
\end{eqnarray}
For a medium with a drift component perpendicular to the strata (i.e. $v_0\neq 0$) 
and a point-like injection (i.e. $L=0$) , we get 
\begin{eqnarray}
  \lefteqn{(\delta D(t))^2
    =\int\limits_{k_1}\int\limits_{k_1} C(k_1)C(k_2)}\nonumber\\
  & & \quad \left[ {\rm e}^{-2k_1^2 D_2 t} \quad F(k_1,k_2)F^{*}(k_1,k_2)\right.\nonumber\\
  & & \quad +\left. {\rm e}^{-{\rm i}v_0(k_1-k_2)t} {\rm e}^{-(k_1^2 + k_2^2)D_2 t} \quad
    F(k_1,k_2)F^{*}(k_2,k_1)\right]
\end{eqnarray}
where $C(k)$ is the partially integrated correlation function
\begin{eqnarray}
C(k) := \int_{k'_1}\int_{k'_2}\int_{k_1} \overline{\tilde{u}(\mathbf{k})\tilde{u}(\mathbf{k'})} = \sqrt{2\pi}\sigma_0^2 l \exp{\left(-\frac{k^2l^2}{2}\right)}\label{corr_fn}
\end{eqnarray}
and
\begin{eqnarray}
  \lefteqn{F(k_1,k_2)=}\nonumber\\
  & & \frac{1}{{\rm i}v_0
    k_2+k_2^2 D_2}\left({\rm e}^{-({\rm i}v_0k_2+k_2^2D_2)t}-1\right)\nonumber\\
  & - & \frac{1}{{\rm i}v_0 k_2+k_2^2 D_2 + 2 k_1 k_2 D_2}\left({\rm e}^{-({\rm i}v_0
      k_2+k_2^2D_2+ 2 k_1 k_2 D_2)t}-1\right) \quad .
\end{eqnarray}
Rescaling the integration variable $k_i=q_i / \sqrt{t}$ and
expanding the resulting expression to $O(1/\sqrt{t})$ gives
\begin{eqnarray}
  \lefteqn{F(q_1/\sqrt{t},q_2/\sqrt{t}) =}
  \nonumber\\
  & & \frac{1}{v_0 q_2} \left[\left(1 - \frac{q_2 D_2}{i
        v_0\sqrt{t}}\right)\left({\rm e}^{-({\rm i}v_0q_2\sqrt{t}+q_2^2D_2)}-1\right)\right.
  \nonumber\\
  & -& \left.\left(1 - \frac{q_2 D_2}{{\rm i}v_0\sqrt{t}} - \frac{2 q_1
        D_2}{{\rm i}v_0\sqrt{t}}\right)\left({\rm e}^{-({\rm i}v_0q_2\sqrt{t}+q_2^2D_2+
        2 q_1 q_2 D_2)}-1\right)\right]
\end{eqnarray}
The limit $t\to\infty$ leads to the expression
\begin{eqnarray}
\lefteqn{(\delta D(t) )^2 =\frac{1}{u_2}\int\limits_{q_1} \int\limits_{q_2} C(0)C(0)\left(\frac{1}{q_1 q_2} + \frac{1}{q_2^2}\right)\times}\label{4.31}\\
& & \times {\rm e}^{(q_1^2+q_2^2)D_2}{\rm e}^{(q_1+q_2)^2D_2}\left({\rm e}^{-q_1q_2 D_2} - {\rm e}^{q_1q_2 D_2}\right)^2+\; O\left(\frac{1}{\sqrt{t}}\right)\nonumber
\end{eqnarray}
The case of finite $L$ can be treated by replacing $F(q_1/\sqrt{t},q_2/\sqrt{t})$ by
\begin{eqnarray}
  \lefteqn{G(q_1/\sqrt{t},q_2/\sqrt{t}) =}\\
  & & \frac{1}{{\rm i}v_0q_2
    +q_2^2D_2/\sqrt{t}}\left({\rm e}^{-({\rm i}v_0q_2\sqrt{t}+q_2^2D_2)}-1\right)
  \nonumber\\
  & - & \frac{{\rm e}^{-q_1 q_2 L^2/t}}{{\rm i}v_0q_2 +q_2^2D_2/\sqrt{t}+ 2 q_1
    q_2 D_2/\sqrt{t}}\left({\rm e}^{-({\rm i}v_0q_2\sqrt{t}+q_2^2D_2+ 2 q_1 q_2
      D_2)}-1\right) \nonumber
\end{eqnarray}

\newpage

\end{document}